\documentclass[superscriptaddress,showkeys,amsmath,amssymb,prl,twocolumn,floatfix,aps]{revtex4}

\usepackage{graphicx}

\begin{document}
\title{Simulation of magnetoresistance in disordered ultracold atomic Bose gases}
\date{\today}
\author{J. Towers}
\affiliation{The Jack Dodd Centre for Quantum Technology, Department of Physics, University of Otago, Dunedin, New Zealand}
\affiliation{Centre for Quantum Technologies, National University of Singapore, 3 Science Drive 2, Singapore 117543}
\author{S.~C. Cormack}
\affiliation{The Jack Dodd Centre for Quantum Technology, Department of Physics, University of Otago, Dunedin, New Zealand}
\author{D.~A.~W. Hutchinson}
\affiliation{The Jack Dodd Centre for Quantum Technology, Department of Physics, University of Otago, Dunedin, New Zealand}
\affiliation{Centre for Quantum Technologies, National University of Singapore, 3 Science Drive 2, Singapore 117543}

\begin{abstract}
Anderson localization was first investigated in the context of electrons in solids. One of the successes was in explaining the puzzle of negative magneto-resistance - as early as the 1940s it had been observed that electron diffusion rates in some materials can increase with the application of a magnetic field. Anderson localization has now been demonstrated in ultra-cold atomic gases. We present a theoretical study of the two-dimensional ultra-cold Bose gas in the presence of disorder, to which we apply a synthetic magnetic field. We demonstrate that, in the ballistic transport regime this leads to positive magneto-resistance and that, in the diffusive and strong localization regimes, can also lead to negative magneto-resistance. We propose experimental scenarios to observe these effects.
\end{abstract}

\maketitle

The study of disorder induced localization in ultracold atomic gases is now well established with strong localization observed in one-dimensional (1d) quasi-periodic lattices\cite{Inguscio} and Anderson localization\cite{Anderson,RevAnderson} in both one\cite{Billy} and three dimensional (3d) geometries\cite{Bouyer3D} with disorder induced through laser speckle\cite{Marie, RevSanchez}. In general, localization is always expected in a disordered 1d system, whereas in 3d there exists a mobility edge\cite{mobility} between localized and extended states and a quantum phase transition between metallic and insulating phases can be expected \cite{Mott}. Two dimensions (2d), as is often the case, is the marginal dimension between these behaviours and, in the solid state, has lead to interesting debate\cite{Kravchenko} regarding the potential observation of a metalic phase in Si MOSFETs. The observation of Anderson localization in 2d in ultracold gases is also complicated\cite{Bouyer_private} by the possibility of impurity potentials leading to classical trapping of the gas when the intensity of the speckle is sufficient to induce localization. Studies of localization in disordered ultracold 2d gases are therefore of timely interest.

Anderson localization is a single-particle interference phenomenon and is strongly enhanced in 2d by the increased occurance of crossing trajectories (in a path integral picture) over and above 3d. Crossing paths always result in closed loops that constructively interfere back at the origin of the path with their time-reversed equivalent. This enhances the probability that a particle that starts at point {\bf r} remains at point {\bf r}, i.e. is localized. If one introduces a magnetic field (of any orientation in 3d, or with some component perpendicular to the plane in 2d) then this time-reversal symmetry is broken and the enhancement of localization destroyed. This is the origin of {\em negative magneto-resistance}, which was a thirty year puzzle until explained in the context of Anderson localization\cite{Lee} in 1980. The observation of the analogue of such negative magneto-resistance in an ultracold atomic gas localized by disorder would be unambiguous evidence that the localization was an interference phenomenon and not classical trapping or interaction induced self-trapping.

Of course, ultracold atoms are charge neutral, so we cannot simply impose an external magnetic field to break the time-reversal sysmmetry. We can however introduce a synthetic magnetic field by rapidly rotating the system \cite{Dalibard} or, more practically, through the use of spatially dependent light fields to couple between internal states of the atoms\cite{Dalibard_synth, Jaksch, Mueller}.

In this Letter we examine a 2d ultracold Bose gas in an optical lattice with quasi-periodic disorder induced by a weak second lattice of incommensurate wavelength to the first. We then impose a synthetic gauge field based upon the Raman scheme\cite{Spielman} to simulate an applied magnetic field, breaking time-reversal symmetry. We demonstrate that, in the ballistic transport regime this leads to (normal, positive) magneto-resistance and in the diffusive (weak localization) and insulating (strong localization) regimes can induce negative magneto-resistance.

Our system is well described by the Bose-Hubbard Hamiltonian which in the absence of inter-particle interactions takes the form:
\begin{equation} \label{eqn:BH-H}
\hat{H} = -\sum_{\langle n,m \rangle}{J_{n,m} \hat{a}_{n}^{\dagger} \hat{a}_{m}} + \sum_{n}{\epsilon_{n} \hat{a}_{n}^{\dagger} \hat{a}_{n}}
\end{equation}
where $\hat{a}_{n}$ is the Bose annihilation operator for the $n$\textsuperscript{th} site, $J_{n,m}$ represents hopping from the $n$\textsuperscript{th} to the $m$\textsuperscript{th} site and $\langle n,m \rangle$ indicates the nearest neighbours $m$ of the $n$\textsuperscript{th} site\cite{Morsch,Wannier}.

Disorder is introduced via interference with a weak optical lattice that is incommensurate with the primary. This, along with an external harmonic trap, is included in the on-site energy term:
\begin{equation}
\begin{split}
\epsilon_{n} = V_{\rm{dis}} \left[\cos(4 \pi x_{n}/\lambda_{2} + \phi_{x}) + \cos(4 \pi y_{n}/\lambda_{2} + \phi_{y}) \right]\phantom{0}\\ + V_{\rm{trap}} (x_{n}^2+y_{n}^2)/\lambda_{1}^2,
\end{split}
\nonumber
\end{equation}
where $V_{\rm{dis}}$ represents the strength of the secondary lattice, $V_{\rm{trap}}$ the strength of the harmonic confinement and $\lambda_{1}$ ($\lambda_{2}$) is the wavelength of the primary (secondary) lattice.

To include the effect of the synthetic gauge field we first assume that there still exists an orthonormal set of basis states that are localized to each site (modified Wannier functions), then impose gauge invariance for any observable on the Hamiltonian\cite{Graf,Luttinger,Boykin}. This motivates the modification to the hopping term in (\ref{eqn:BH-H}):
\begin{equation}
J_{n,m} \rightarrow J_{n,m}^{({\bf A})}  = J_{n,m} \exp\left[-\frac{i q}{\hbar} \int_{{\bf x}_{n}}^{{\bf x}_{m}}{{\bf A} \cdot {\bf ds}}\right],
\end{equation}
where the integral is taken over the shortest path separating the end points. With this it can be shown that any gauge transformation with a function $f(x,y)$, such that $A \rightarrow A' = A + \nabla f$, results in $\hat{a}_{n} \rightarrow \hat{a}_{n}' = \exp\left[-i q f/\hbar\right] \hat{a}_{n}$ and conserves the site number operator ($\hat{n}_{n}' = |\hat{a}_{n}'|^2 = |\hat{a}_{n}|^2 = \hat{n}_{n}$), therefore satisfying gauge invariance.

The spatial dependence of the hopping terms can be calculated analytically, if one approximates the Wannier functions of the tight binding model\cite{Bullett} with Gaussians, or numerically with the (appropriately modifed, in the presence of a field) Wannier functions themselves. In the absence of a magnetic field it is our experience\cite{clemens}, and is well established in the literature\cite{Guarrera, Schmitt}, that the spatial dependence of the hopping term is of negligible importance compared to the on-site energy fluctuations. We therefore assume that this remains true in the presence of a magnetic field and set $J_{n,m} = J$.

We then apply the Mean-Field approximation ($a_{n} \simeq \langle a_{n} \rangle = z_{n} : z_{n} \in \mathbb{C}$) to the Heisenberg equation of motion for the Bose annihilation operator. It is convenient to relabel the sites according to their spatial position, we therefore define $z_{j,k}$ to be the amplitude of the site located at $(x,y) = (k, j) a_{1}$ : $j,k \in \mathbb{Z}$, $a_1 = \lambda_{1}/2$. The discrete mean-field equations of motion in the Landau gauge (${\bf A} = -B y {\bf \hat{x}}$) are then given by:
\begin{equation}
\begin{split}
i \dot{z}_{j,k} = -z_{j-1,k} -z_{j+1,k}\phantom{0000000000000000000}\\ -e^{-i \varphi j} z_{j,k-1} -e^{i \varphi j} z_{j,k+1} + \epsilon_{j,k} z_{j,k}
\end{split}
\end{equation}
with:
\begin{equation}
\begin{split}
\epsilon_{j,k} = \Delta \left[\cos(2 \pi \alpha j + \phi_{y}) + \cos(2 \pi \alpha k + \phi_{x}) \right] \phantom{00}\\+ v_{\rm{trap}} (j^2+k^2),
\end{split}
\end{equation}
where the dot indicates the derivative with respect to scaled time $\tau = t/t_0$, $t_0 = J/\hbar$. $\Delta = V_{\rm{dis}} / J$ is the secondary lattice parameter and $v_{\rm{trap}} = V_{\rm{trap}} / J$ the harmonic trap parameter. $\varphi$ is the phase accumulated in the Aharanov-Bohm effect when a charged particle circumnavigates one lattice plaquette in the anti-clockwise direction.

If we neglect the synthetic gauge field this system is separable into two 1d systems, each of which, if we set the lattice wavelength ratio $\alpha$ to an irrational number, is equivalent to the Aubry-Andr{\'e} model\cite{AA}. Therefore, for $\Delta < 2$ the eigenstates are extended and the dynamics are ballistic; for $\Delta = 2$ the dynamics are diffusive; and for $\Delta > 2$ the eigenstates are exponentially localized and we enter the Anderson Localized regime where, except for virtual transmission on the order of the localization length, conduction is entirely suppressed\cite{Larcher}. In the presence of the synthetic gauge field, however, the equations are not separable and we find some interesting dynamics.

\begin{figure}[!ht]
\includegraphics[width=0.4\textwidth]{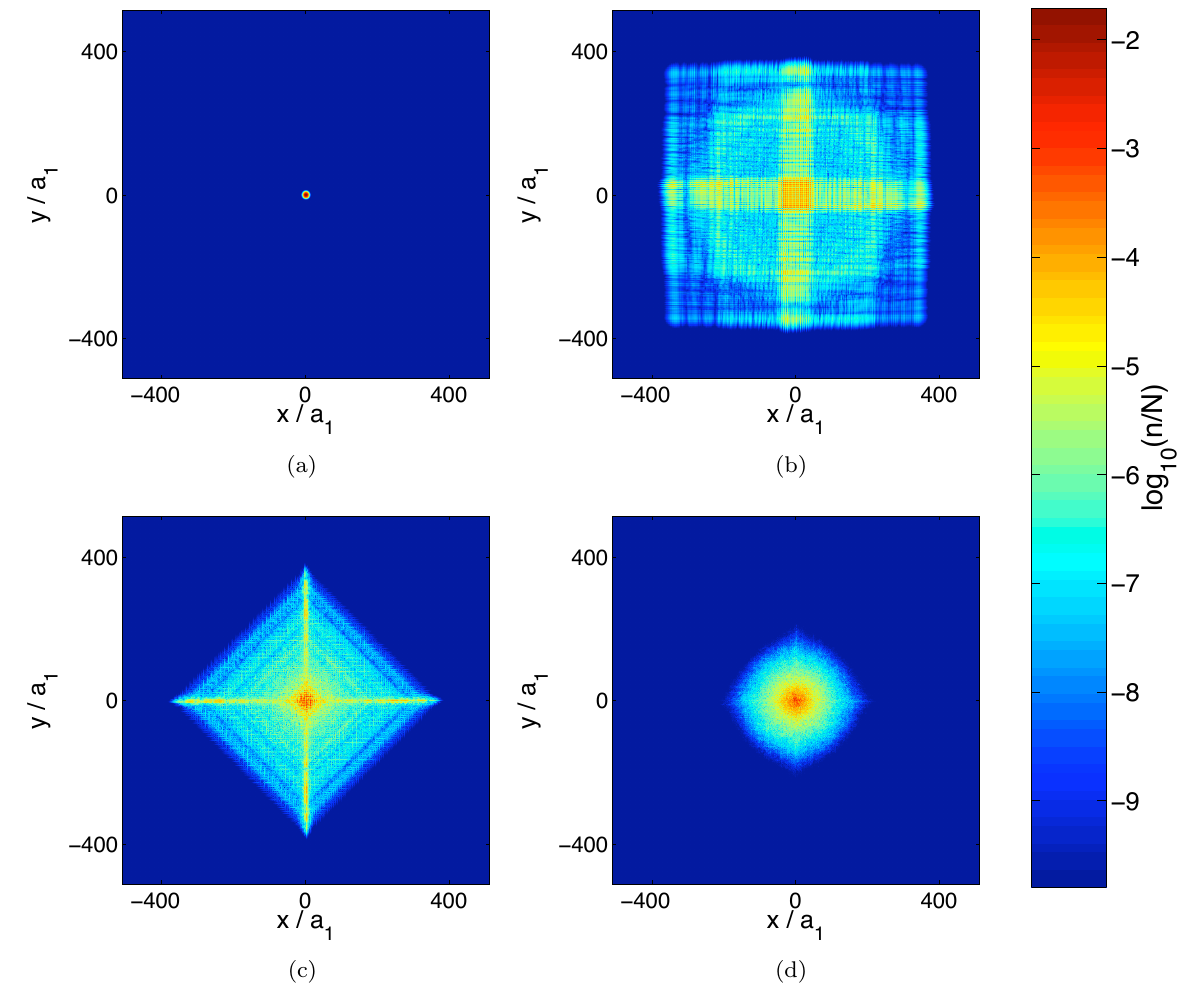}
\caption{Site occupancy (logarithmic) of (a): the initial wave-packet at $t=0$ and (b-c): after propagating for $1000 t_0$s in the secondary lattice, with $\Delta = 1$, and synthetic gauge field (b) $\varphi=0.001$; (c) $\varphi=0.1$; (d) $\varphi=1$.\label{fig:panels}}
\end{figure}

For all simulations we initialize the wave-packet in the ground state of the primary lattice and a fairly strong harmonic trap ($v_{\rm{trap}} = 10^{-2}$). This ground-state has radial symmetry in a nearly Gaussian distribution [Fig. \ref{fig:panels} (a)]. The harmonic trap is then switched off at the same time as the synthetic gauge field and secondary lattice potential are switched on. The changing magnetic field gives the wave-packet a gauge dependent momentum kick\cite{SynthE}. In the symmetric gauge the wave-function remains unchanged as the contributions from the two dimensions cancel. We work in the Landau gauge, however, so we must transform the wave-function using $z^{L}_{j,k} = e^{i \varphi j k / 2} z^{S}_{j,k}$.

The periodicity of the primary lattice introduces a cosine dispersion relation in the first Brillouin zone. The lowest momentum states fall on the approximately quadratic part of the dispersion relation and so the motion of an initially low energy wave-packet mimics that of a wave-packet in free space but with a reduced effective mass\cite{A+M}. The motion is termed ballistic as the {\it rms}. half-width scales as $t$ with time.

In this picture, we can understand our system in the presence of a synthetic gauge field. A classical particle in a magnetic field would complete closed circular loops in the 2D plane. In our simulations, an initially Gaussian distribution expands to a larger Gaussian with a ragged boundary that appears to be rotating. The size to which the Gaussian can expand is determined by the starting size and the size of the classical cyclotron orbits. So far, this system is analogous to that described by \cite{Hofstadter} which results in the celebrated Hofstadter butterfly, also investigated recently\cite{Powell} in the context of atomic gases.

\begin{figure}[!ht]
\includegraphics[width=0.4\textwidth]{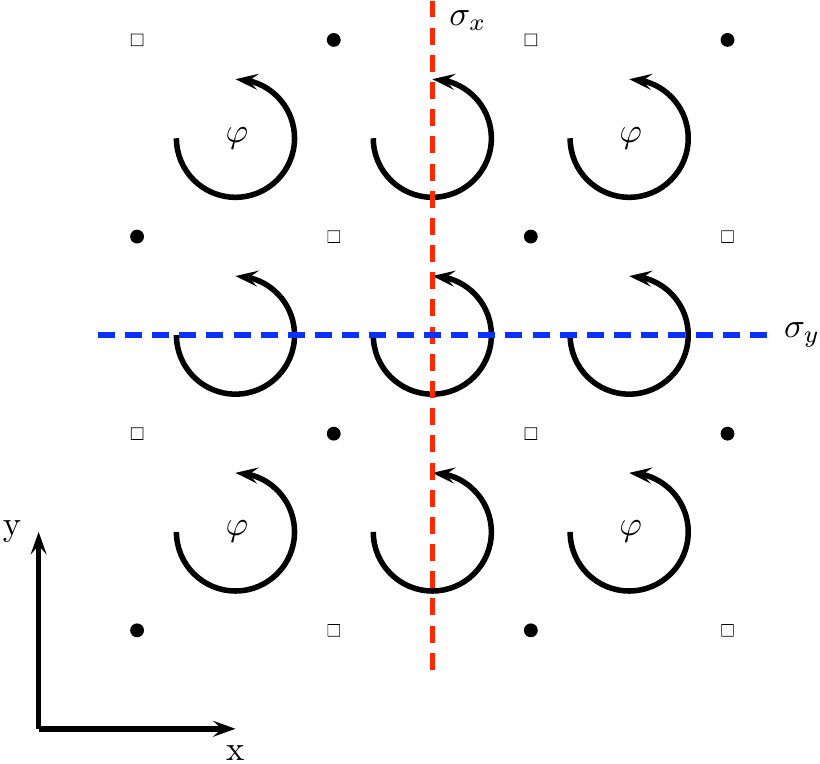}
\caption{Diagram of the primary lattice depicting the phase accumulated around a lattice plaquette in the synthetic gauge field and the symmetries, $\sigma_x$ and $\sigma_y$, which are broken. Filled and unfilled dots indicate the loss of self-similarity between these sets of sites due to symmetry breaking.\label{fig:symmetry}}
\end{figure}
\begin{figure}[!ht]
\includegraphics[width=0.4\textwidth]{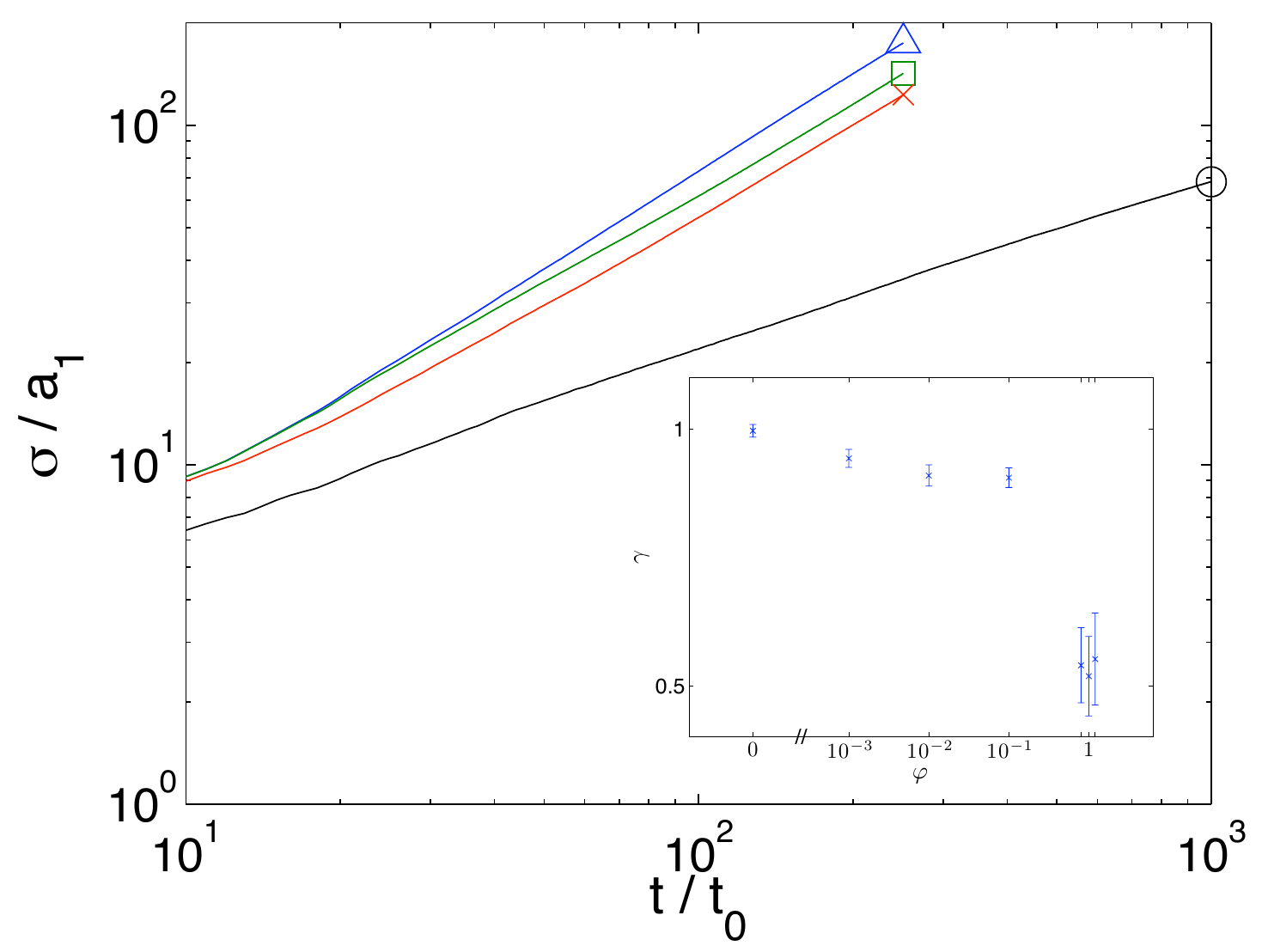}
\caption{{\it Rms}. half-width of the wave-packet vs. time during evolution in a secondary lattice with strength $\Delta = 1$ and synthetic gauge field: (triangle) $\varphi = 0.001$; (square) $\varphi = 0.01$; (cross) $\varphi = 0.1$; (circle) $\varphi = 1$.\\
(Inset): Long term scaling ($\sigma \propto t^{\gamma}$) of the width of the wave-packet.\label{fig:D1}}
\end{figure}
In the extended regime ($\Delta < 2$), the wave-packet released in to a negligibly weak magnetic field gains the square symmetry\cite{fermHubb} of the reciprocal lattice, with the real space density at long times reflecting the initial momentum distribution, [Fig. \ref{fig:panels} (b)] and the {\it rms. }half-width follows the expected $t$ dependence [Fig. \ref{fig:D1} - triangle] characteristic of ballistic expansion.

When the magnetic field is stronger, time-reversal symmetry is broken and hence the $\sigma_x$ and $\sigma_y$ symmetries\cite{Joshua} (reflection in the $x=0$ and $y=0$ planes respectively) are also broken. The $C_2 = \sigma_x \sigma_y$ symmetry (rotation by $\pi$) remains, however, and so the solid points and unfilled squares in Fig. \ref{fig:symmetry} are each self-similar and the Bravais lattice and hence the reciprocal lattice is now rotated by $\pi/2$. The lattice parameter is also increased by a factor of $\sqrt{2}$ and hence the `volume' of the Brillouin zone is reduced by half. This results in the diamond symmetry of Fig. \ref{fig:panels} (c), a signature that would be clearly observable in time-of-flight type experiments.

In a very strong gauge field [Fig. \ref{fig:panels} (d)], the initial momentum distribution no longer fills the entire Brillouin zone and so the expanded wave-packet maintains some of the radial symmetry, losing the diamond-like structure.

At long times, increasing the magnetic field has the effect of decreasing the wave-packet's rate of expansion [Fig. \ref{fig:D1}]. The classical analogy is that the particles are completing more of their tighter circular orbits, and hence traversing less linear distance, before being scattered off the pseudo-random secondary lattice. This is therefore analogous with normal, positive magneto-restistence.

In the inset of Fig \ref{fig:D1} we plot $\gamma = \frac{d \rm{log}_{10}(\sigma)}{d \rm{log}_{10}(t)}$ at long times for a selection of gauge field strengths. The data points are obtained using a temporal average about each point. The error bars reflect the variance in the data. The long time behaviour of the size of the wave-packet approaches $t^{1/2}$ with very strong magnetic field, which is characteristic of diffusive expansion.

To summarize, for weak disorder, the effect of the magnetic field is therefore to change the transport from ballistic to diffusive expansion. Furthermore, with increasing field strength, the diffusion coefficient is reduced, consistent with positive magneto-resistance.

\begin{figure}[!ht]
\includegraphics[width=0.4\textwidth]{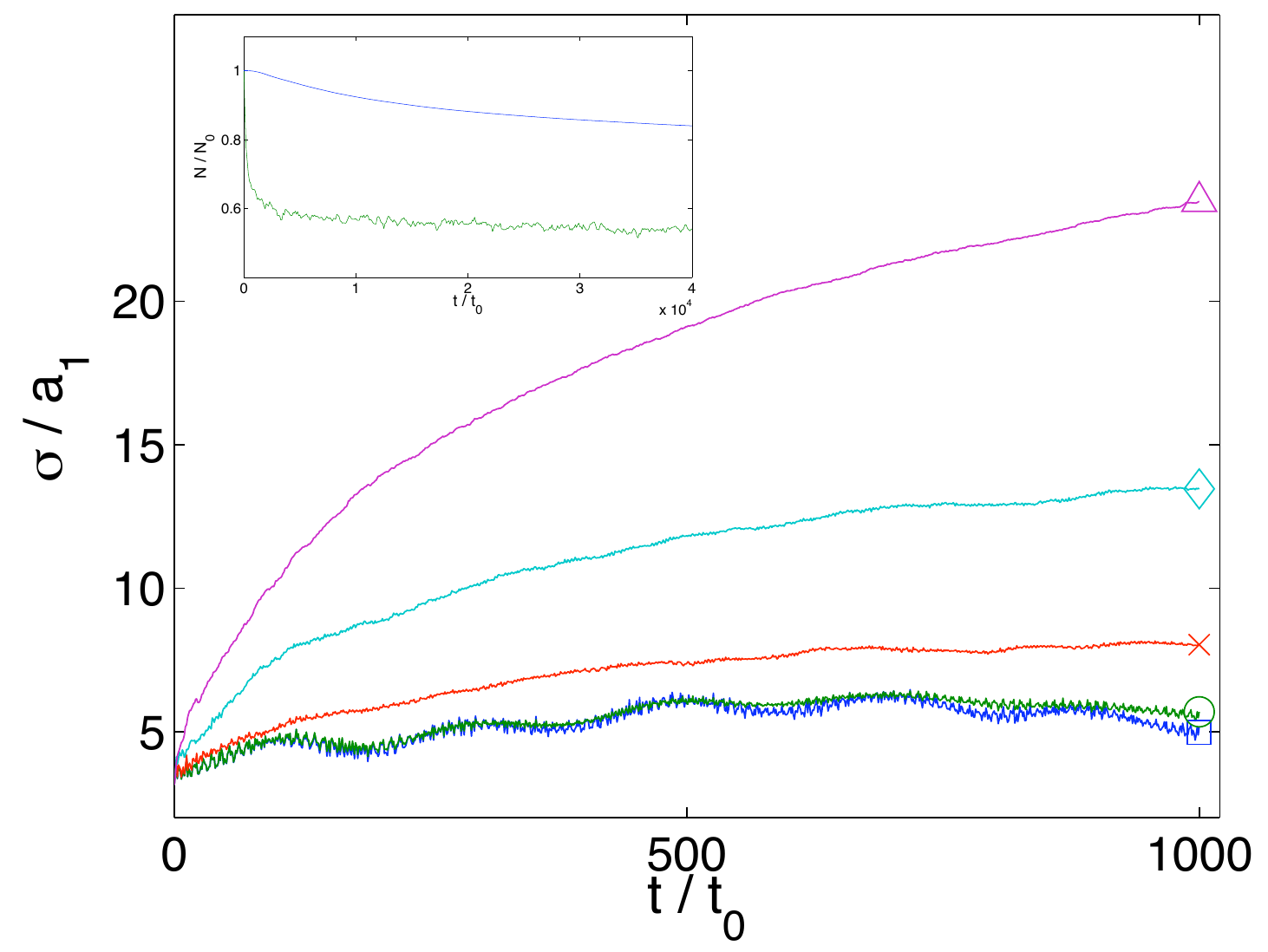}
\caption{{\it Rms}. half-width of the particle wave-packet vs. time during evolution in a secondary lattice with strength $\Delta = 3$ and synthetic gauge field: (square) $\varphi = 0.001$; (circle) $\varphi = 0.01$; (cross) $\varphi = 0.1$; (diamond) $\varphi = 0.5$; (triangle) $\varphi = 1$.\label{fig:D3}\\
(Inset): Total particles vs. time when $\Delta = 3$, $\varphi = 1$ and particles in the outer 25 sites of the $256\times256$ system are strongly attenuated. (dashed line) Full system, (solid line) $31\times31$ sites at center of the system.}
\end{figure}
In the Strongly Localized regime, $\Delta > 2$, we observe the confinement of the wave-packet consistent with the Aubry-Andr{\'e} model for a sufficiently weak gauge field [Fig. \ref{fig:D3} - square]. This Anderson Localization is the result of the total destructive interference of multiply scattered matter-waves for any sites beyond the localization length. 

Considering any path that loops back to its origin, the same path traversed in the opposite (time-reversed) direction is of the same length and hence will return the same phase. Any closed path and its time-reversed partner will therefore interfere constructively. In the presence of a gauge field, however, the phase is displaced proportional to the flux enclosed which has opposite sign for each direction (the Aharonov-Bohm effect). The two paths will then interfere with an essentially random phase. When averaged over many paths, the backscattering of the matter-waves is now dramatically reduced, resulting in the destruction of the Anderson localization and a positive expansion of the wave-packet.

For a sufficiently strong gauge field we clearly observe the destruction of the Anderson localization [Fig. \ref{fig:D3}]. Furthermore, increasing the magnetic field increases the rate of expansion of the wave-packet, characteristic of negative magneto-resistance. The transition occurs continuously suggesting that Anderson Localization is broken for any magnetic field strength, although it may be undetectable for the system run times that we can simulate.

It is interesting to note that localization is not completely broken for all states. Some of the eigen-states remain localized. In the inset of Fig. \ref{fig:D3} we demonstrate the presence of localized eigen-states. For this plot we have included a large negative-imaginary term in the local part of the Hamiltonian for sites within 25 sites of the system edge (in a $256\times256$ site system). When released from the harmonic trap, any outward bound population is quickly attenuated before it can create edge effects which reflect back to affect the center. In the plot we observe that the decay of the total population saturates as all the extended states are attenuated when they reach the edge of the simulation grid, leaving behind the surviving localized states. This is most pronounced for the $31\times31$ sites at the center of the system. Such behaviour could be observed experimentally by taking in situ absorbtion images and observing the change in particle number with time.

In conclusion, we have used numerical simulations to demonstrate the interplay between synthetic gauge fields and Anderson localization in the Aubrey-Andr{\'e} model. We observe firstly positive magneto resistance in the extended regime and then negative magneto-resistance in the strong localization regime. We have demonstrated distinctive behaviours for each regime that should be experimentally observable through straightforward absorption imaging techniques. Especially, the observation of negative magneto-resistance can only be explained in the context of an interference phenomenon. This would therefore be an unambiguous signature that localization, destroyed or reduced by the imposition of a magnetic field, had such interference as its origin, distinguishing it from classical trapping or interaction-induced self-trapping. This is especially important in 2D where ambiguity in the origin of localization in experiments still resides.

We would like to acknowledge financial support from the NZ Foundation for Research, Science and Technology through contract NERF-UOOX0703 and from the National Research Foundation and Ministry of Education of Singapore.

\end{document}